\documentstyle[psfig]{mn}
\pssilent

\def \APJ { ApJ}
\def \MN  { MNRAS}

\def\Mpc{\, h^{-1} \, {\rm Mpc}}
\def \etal {et al.}
\def \eg { {\it e.g.} }

\def \degr { ^{\circ}}

\begin{document}

\title[Large scale structure in deep radio surveys]{Searching for large scale structure in deep radio surveys}

\author[A.~Baleisis, O.~Lahav, A.~J.~Loan, and J.~V.~Wall]{Audra Baleisis,$^{1,3}$,
Ofer Lahav,$^{1,4}$, Andrew J.~Loan,$^1$ and Jasper V.~Wall$^2$\\ 
$^1$ Institute of Astronomy, Madingley Road, Cambridge, CB3 0HA \\ 
$^2$ Royal Greenwich Observatory, Madingley Road, Cambridge, CB3 0EZ \\ 
$3$ University of Arizona, Steward Observatory, Tucson, AZ 85721, U.S.A. \\
$4$ Anglo-Australian Observatory, P.O. Box 296, Epping, NSW 2121, Australia}

\maketitle

\begin{abstract}
  We calculate the expected amplitude of the dipole and higher
  spherical harmonics in the angular distribution of radio galaxies.  The
  median redshift of radio sources in existing catalogues is $z \sim
  1$, which allows us to study large scale structure on scales between
  those accessible to present optical and infrared surveys, and that of the
  Cosmic Microwave Background (CMB).  
  The dipole is due to 2 effects which turn out to be of
  comparable magnitude: (i) our motion with respect to the 
  CMB, and (ii) large scale structure, 
  parameterised here by a family of Cold Dark Matter power-spectra.  We
  make specific predictions for the Green Bank 1987 (87GB) and Parkes-MIT-NRAO (PMN) catalogues, which
  in our combined catalogue 
  include $\sim 40,000$ sources brighter than 50 mJy at 4.85 GHz,
  over about 70\% of
  the sky.  For these relatively sparse catalogues both the motion and
  large scale structure dipole effects are expected to be smaller than the
  Poisson shot-noise.  
  However, we detect dipole and higher harmonics in the combined ${87GB-PMN}_{raw}$ 
  catalogue which are far larger than expected.  We
  attribute this to a 2\% 
  flux mismatch between the two catalogues. 
  Ad-hoc corrections to match the catalogues may suggest a marginal
  detection of a dipole.
  To detect a dipole and higher harmonics unambiguously,
  a catalogue with full sky coverage and $\sim 10^6$ sources is
  required.  
  We also investigate the existence and extent of the Supergalactic Plane in 
  the above catalogues. In a strip of $\pm 10^\circ$ 
  of the standard Supergalactic equator,  
  we find a $3\sigma$ detection in ${PMN}_{raw}$, 
  but only $1\sigma$ in ${87GB}_{raw}$.
  We briefly discuss the implications of  on-going surveys such
  as FIRST and NVSS and follow up redshift surveys.
\end{abstract}

\section{Introduction}

Our local universe is overwhelmingly clumpy.
As we look on
increasingly larger scales, we continue to see clustering. Observations of discrete objects and their
clustering properties on $\sim100 \Mpc$ scales start to show evidence for
homogeneity. The largest scales we observe ($\sim 1000 \Mpc$) are by
looking at the Cosmic Microwave Background (CMB).  In the CMB (apart from
a dipole component attributed to our motion) we see a striking level 
of isotropy, with fluctuations at the level of 
$10^{-5}$ on $10\degr$ scales (Smoot et al. 1992).
Somewhere between the scale accessible to us by the CMB and
that of galaxies and quasars, the universe changed from smooth to lumpy,
forming structures from some small initial fluctuations.

It has been suggested (Kaiser 1984) that galaxies form preferentially in
high density peaks of the underlying mass distribution. If this is true,
then the statistics of galaxy distributions provide us with information,
albeit biased, about the underlying dark matter distribution.

Optical and infrared galaxy surveys have been used extensively 
to study clustering out to  $\sim 200 \Mpc$, or a redshift of $z \sim .07$ (e.g. Strauss \& Willick 
1995 for review).
The high luminosity of
radio galaxies and quasars, makes them detectable out to large,
cosmological distances ($z \sim 4$)
and consequently, useful in studying clustering out
to high redshift. Unfortunately, most large surveys to date provide only
fluxes and angular coordinates on the sky, not redshifts, for individual
objects. This means that at present, only the projected distributions of
radio galaxies can be studied.

It has long been debated whether radio galaxies are clustered or isotropic
on the largest scales.  The much quoted study by Webster (1976), which
looked at 8000 radio sources at 408 MHz, found $<3\%$ variability in the
number of sources found in a randomly placed 1 Gpc cube. This led to the
generally accepted view that radio sources were isotropically distributed.
Even if this were not true, the large range in intrinsic luminosities of
radio sources would effectively wash out structures when the
distribution was projected onto the sky and information about the objects' radial distribution was lost.  More recently, Shaver \& Pierre
(1989) reported a detection of slight clustering of bright, nearby radio
sources to the Super Galactic Plane (SGP) and Benn \& Wall (1995) 
discussed other measures of anisotropy in radio surveys.  In Section 7
we investigate the presence of the SGP in the ${87GB}_{raw}$ and ${PMN}_{raw}$ and
report on marginal detection.
Clustering in the  above catalogues was also studied 
by correlation function analysis (Kooiman, Burns \& Klypin 1995;
Sicotte 1995; Loan, Wall \& Lahav 1997). These studies indicated 
that radio galaxies are actually more strongly clustered than 
local optical galaxies. This conclusion is also confirmed by 
correlation analysis of the FIRST survey (Cress et al. 1997).
Motivated by these recent indicators of clustering in 
radio surveys and by the upcoming deeper surveys 
such as FIRST and  NVSS
(see Condon 1997 for review), we attempt here to study large scale
structure in radio surveys using Spherical Harmonic Analysis
(SHA).    

At this point, we introduce the labels that we use to refer to various forms of the 87GB and PMN catalogues. From here on, we will distinguish between the raw, uncorrected, combined catalogue (${87GB-PMN}_{raw}$), and our $\cal N$ matched (corrected) version of the combined catalogue (${87GB-PMN}_{match}$). When talking about predictions, we use the label 87GB-PMN (with no subscripts) to refer to an ideal 4.85 GHz survey. 

In Section 2, we
present the ${87GB}_{raw}$ and ${PMN}_{raw}$ catalogues 
and discuss their mismatch. 
In Section 3, we show the formalism 
of predicted rms harmonics for the models.  
In Section 4, we look at the radio
dipole and how it compares to the predicted dipoles due to shot noise, our
motion with respect to the CMB, and large scale structure. 
In Section 5, we
consider the observed dipole in the combined catalogue, 
and in Section 6, apply the SHA to the catalogue.
In Section 7, we discuss the Supergalactic Plane
and in Section 8, we summarize the results and discuss future surveys.  

\section{The Radio catalogues}
\label{radio}

\subsection{The Green Bank survey (87GB)}

The 87GB survey was carried out by Condon, Broderick \& Seielstad (1989),
over a period of three years (1986-88), using the NRAO 91m transit radio
telescope (in Green Bank, West Virginia, USA) in conjunction with the NRAO
4.85 GHz, 7-beam receiver that was built for the survey. A large part of
the sky visible from the Northern Hemisphere was observed ( $0\degr <
\delta < +75\degr$, $0^h < \alpha < 24^h$). Errors due to instrument noise
and source confusion for the survey were estimated to be $\sigma \sim 5$mJy and a detection threshold of $S \ge 5\sigma$ was adopted in compiling
the source lists.  A catalogue of 54,579 radio sources above a flux limit
of $\sim 25$mJy was subsequently produced (Gregory \& Condon, 1991) from
observations taken in 1987.  Excluding a few small regions thought to be
contaminated by sidelobes of strong local sources or by solar interference,
data from all the surveyed regions were included in the source catalogue,
which covers $\sim 6$ steradians on the sky.

\subsection{The Parkes-MIT-NRAO survey (PMN)}

The PMN survey was carried out by Griffith \& Wright (1993) over the course
of 1990, using the Parkes 64m telescope (near Parkes, New South Wales,
Australia) in conjunction with the same 7-beam receiver as used in 87GB
survey. The area to be surveyed, which covered $-87.5\degr < \delta <
+10.0\degr$, for all right ascensions ($0^h < \alpha < 24^h$) in the
Southern Celestial Hemisphere, was divided and observed in 4 declination
strips (see Table 1, Griffith \& Wright 1993). The source detection
reliability criterion was 90\%, corresponding to about $S \ge 4.4\sigma$.
The definition adopted of a real source, was that its observed flux was
close to the flux limit - not necessarily above the flux limit.  Source
catalogues for three of the four declination strips have been complied
(Tropical - by Griffith et al. 1994; Southern - by Wright et al. 1994 and
independently by Gregory et al. 1994; Equatorial - by Griffith et al.
1995), together covering $\sim 6.4$ steradians of sky. More recently, the
source catalogue for the fourth declination strip (Zenith) has become
available (Wright \etal 1996), but because of its lower flux limit of about
72mJy, it is not included in this study.

\subsection{How well are the resulting catalogues matched?}

We intend to use the ${87GB}_{raw}$ and ${PMN}_{raw}$ catalogues together to provide extensive
sky coverage, and need some idea of what inconsistencies might arise from
the fact that different telescopes were used to collect the data and
different reduction algorithms and selection criteria were used in
compiling the catalogues.  Fortunately, there are two regions of the sky
for which there is overlap information.

\subsubsection{Region 1}

The declination band $0\degr<\delta<10\degr$ was observed with both the
NRAO and Parkes telescopes and the data collected reduced by 87GB catalogue
method (Gregory \& Condon 1991; Gregory et. al 1994) as well as the PMN
catalogue method (Griffith et al 1994; Griffith et al. 1995). Griffith
\etal (1995) examine the agreement between resulting catalogues. They find
some discrepancies in positions and a 2\% disagreement of flux scales. A
number of explanations for these inconsistencies are offered:
\begin{itemize}
\item{Two different telescopes were used to collect the data.}
\item{The observations were carried out at 2 different epochs.}
\item{Different reduction algorithms were used.}
\item{Different source detection thresholds were used.}
\end{itemize}

\subsubsection{Region 2 }    

Two catalogues (Gregory et al. 1994; Wright et al. 1994) were produced with
data from the PMN survey, Southern declination strip.  Gregory et al.(1994)
compare the results of two data reduction techniques used to compile the
two source lists:
\begin{itemize}
\item{They find good agreement of positions of strong sources, but up to $1\sigma$ disagreement in faint source positions.}
\item{They find the Wright \etal (1994) resulting catalogue contains more
sources overall.}
\end{itemize}

\subsubsection{Problems within individual  catalogues}

Two effects cause the flux limit of the 87GB survey to vary with
declination, increasing from $\sim$25\,mJy at $\delta > 60\degr$ to
$\sim$40\,mJy at $\delta = 0\degr$ (Gregory \& Condon 1991). One is the
elevation dependence of the sensitivity of the 91\,m NRAO telescope, caused
by a higher noise level at larger angles from the local zenith
($\delta=38\degr$). The other is the fact that adjacent observing tracks
are spaced slightly further apart with decreasing declination.  The Parkes
telescope does not suffer decreased sensitivity at low elevations.
However, the declination-dependent track spacing holds for low declinations
($\delta < -70\degr$) and causes the flux limit to decrease from
$\sim$50\,mJy near the zenith to $\sim$40\,mJy near the pole.

\subsubsection{What we use as our catalogue}

\begin{figure*}
\begin{center}
\parbox{17cm}{\psfig{figure=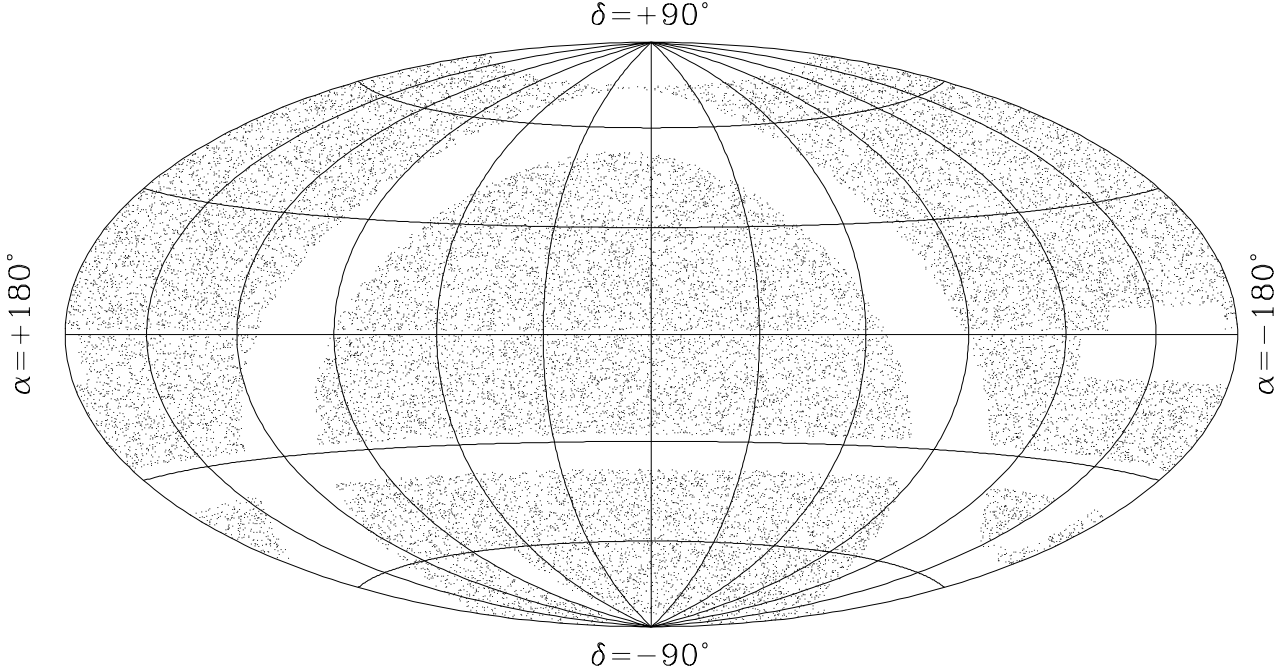,width=17cm,clip=}}
\end{center}
\caption{\label{aitoffdata}The distribution of radio sources (above flux cutoff of 70mJy in
${87GB}_{raw}$, 74 mJy in ${PMN}_{raw}$) from the ${87GB-PMN}_{raw}$ catalogue areas as selected 
by Loan et al. 1997. This is an Aitoff projection
in Equatorial coordinates.}
\end{figure*}

We use the same cuts of the source catalogues as listed in Table 1 in
Loan et al. (1997), resulting 
in sky coverage of $\sim 70 \%$.
We impose a working lower flux limit of 50 mJy.  Figure ~\ref{aitoffdata} shows the distribution of the raw data over the sky, in equatorial coordinates. Here we discuss the procedure used to try correcting 
for the flux and number discrepancies mentioned in the discussion of
the overlapping regions above. A measurement of ${\cal{N}}$ for each catalogue separately, at a given flux
cutoff, shows that ${{\cal{N}}_{PMN-raw} > {\cal{N}}_{87GB-raw}}$ by a few percent
for the range of flux cutoffs considered above as listed in columns 1-4 of Table ~\ref{nmatch}. It is possible
that we are detecting actual large scale structure on the scale of the
catalogues, but it is more likely that the inconsistencies (dicussed in
Section~\ref{radio}) in the methods by which catalogues were compiled are
responsible. 

In an attempt to correct for the incompatibility of the two
catalogues, we chose the following procedure to set ${\cal{N}}_{PMN} \approx
{\cal{N}}_{87GB}$: Mean density ($\cal N$)
matching - We take 87GB fluxes as fiducial. For each flux cutoff (and corresponding $\cal{N}$) in the ${87GB}_{raw}$
catalogue, we find the flux cutoff in the ${PMN}_{raw}$ catalogue that gives
${\cal{N}}_{PMN}$ as close as possible to ${\cal{N}}_{87GB}$. Columns 1, 2,
and 5-7 in Table ~\ref{nmatch} list these values.
The numbers are not perfectly matched 
as the fluxes are quantizd in units of  1mJy. 
The 60 mJy is the worst matched. 

\begin{table*}
\caption{\label{nmatch} Results of mean density ($\cal N$) matching procedure. The
columns labeled \% dif. give the percentage by which the larger $\cal{N}$
exceeds the smaller $\cal{N}$.}
\centerline{
\begin{tabular}{|l|c|c|c||l|c|c|}
\hline \multicolumn{1}{|c|}{$S_{lim}$} &
\multicolumn{1}{c|}{${\cal{N}}_{87GB-raw}$} &
\multicolumn{1}{c|}{${\cal{N}}_{PMN-raw}$} & \multicolumn{1}{|c||}{\% dif.} &
\multicolumn{1}{c|}{$S_{lim}$} & \multicolumn{1}{c|}{${\cal{N}}_{PMN-match}$} &
\multicolumn{1}{c|}{\% dif.} \\ \hline 50 & 4457 & 4929 & 10.5 \% & 54 &
4433 & 0.5 \% \\
%\hline
60 & 3520 & 3836 & 8.9 \% & 61 & 3744 & 6.4 \% \\
%\hline
70 & 2870 & 3123 & 8.8 \% & 74 & 2891 & 0.7 \% \\
80 & 2386 & 2616 & 9.6 \% & 85 & 2403 & 0.7 \% \\
90 & 2021 & 2241 & 10.8 \% & 97 & 2012 & 0.4 \% \\
100 & 1751 & 1922 & 9.8 \% & 107 & 1758 & 0.4 \% \\
200 & 655 & 684 & 4.4 \% & 205 & 658 & 0.5 \% \\
\hline
\end{tabular}
}
\end{table*}
%\end{itemize}

\section{Spherical Harmonic Analysis}
\label{sha}

It is common (e.g. Peebles 1980) to express the 
overdensity at a given point as a superposition of
Fourier modes  $\delta_{\bf k}$:
\begin{equation}
\label{e5}
\delta ({\bf r}) \equiv
{ {\delta \rho} \over \rho } ({\bf r}) =
{1 \over ({2 \pi}) ^3 } \int d^3k \; \delta_{\bf k} \;
 e^{- i {\bf k}\cdot {\bf r}},
\end{equation}
where ${\bf r}$ refers to comoving coordinates.
In linear theory ($\delta({\bf r})\ll1$), the 
perturbation corresponding to a given
wavenumber, $\bf{k}$, is assumed to evolve independently of all other
$\bf{k}$ modes.

An alternative presentation of power on different scales is 
by Spherical Harmonic Analysis (SHA), which has become 
popular in recent years in studying the CMB and galaxy surveys.
The idea of using SHA to study galaxy
distributions was first proposed over 20 years ago (Peebles 1973).  The
successful application of this method to actual data had to wait another
decade for galaxy surveys (e.g. IRAS) that covered enough of the sky and
contained enough objects to make SHA feasible.
Unfortunately, for most of the objects detected, no redshifts were
available.  In this case, the true 3D galaxy distribution is seen projected
onto the surface of the celestial sphere and the spherical harmonic
analysis is done in 2D (Scharf et al. 1992, Lahav 1994). When redshifts
became available, the   spherical harmonic analysis formalism was extended to
3D (Scharf \& Lahav 1993; Fisher, Scharf \& Lahav 1994; Fisher et al. 1995a;
Heavens \& Taylor 1995; Nusser \& Davis 1994). In the 3D case, the redshift
distortion introduced by individual galaxies' peculiar velocities must be
corrected for in order to get distance information. 
As the radio catalogues described here provide only
the angular position of galaxies, we consider here 2D harmonic formalism,
but relate it to the 3D power-spectrum.

\subsection{2D Harmonic Formalism}

Expanding a galaxy distribution in terms of spherical harmonic order, $l$,
gives us information about clustering amplitude as a function of angular
scale, $\Delta \theta \sim {\pi \over l}$, on the sky. This angular scale can be
related to a linear scale by incorporating distance information in the form
of either individual galaxy redshifts or overall redshift distribution,
$N(z)$, of the galaxies.

  The distribution is expanded in spherical harmonics:
\begin{equation}
\label{eqsig}
\sigma({\bf \hat r}) = \sum_{l,m}  a_{l}^{m} Y_{l}^{m}({\bf \hat r}) ,
\end{equation}
where ${\bf \hat r}$ is the angular position on the sky, 
and the resulting harmonic coefficients are 
\begin{equation}
\label{e6}
a_{l}^{m} = \sum_{i=1}^N \; {Y_{l}^{m^*}}({\bf \hat r_i}).
\end{equation}

In the case of incomplete sky coverage 
we derive the harmonic coefficients from the  data as
\begin{equation}
\label{e8}
c_{l,obs}^{m} = {\sum_{i=1}^{N_{gal}} Y_{l}^{m^*}(\bf \hat r_i)} - {\cal{N}} \int_{\Omega_{obs}} d{\Omega} \; Y_{l}^{m^*}(\Omega),
\end{equation}
where $\cal N$ is the number of sources per steradian.
The first term on the right hand side is $a_{l}^{m}$ from eqn.~\ref{e6},
and the second term is the correction for the unsurveyed and excluded
regions of the sky. The second term (calculated by Monte Carlo
integration) is included so that, for a
Poisson distribution of galaxies, $c_{l,obs}^{m} = 0$ (for$ \; l > 0$)
when $N_{gal} \rightarrow \infty$. 
%We calculate the second term on the
%right side of eqn.~\ref{e8} by filling in the observed regions of the sky
%with many
%(e.g. 500,000)
% random points, calculating the harmonic coefficients for those
% points, and then normalising the coefficients to the mean density for the
% flux cutoff we are using.

\subsection{Predicted Harmonics}

We can predict what the above $a_{l}^{m}$ should be, for a given cosmology and power
spectrum of fluctuations, $P(k)$ (dictated by a dark matter model).
Note that the 2D spherical harmonic analysis formalism can be extended to
other cosmologies (Langlois et al, in preparation), 
but for this discussion we assume an Einstein-de Sitter
universe ($\Omega =1, \Lambda = 0$).

We also account for the fact that a galaxy survey will only detect a
fraction of the radio galaxies that populate the surveyed volume by
including the selection function, $\Phi{(r)}$, which is the
probability that a radio galaxy will be detected at distance $r$ in
this flux limited survey.  The observed galaxy distribution is
expanded in summation because galaxies are discrete objects. To model
the underlying mass distribution, which is continuous, the expansion
is in integral form. Here we follow the formalism developed by Lahav,
Piran \& Treyer (1997) for the X-ray Background. As the sources of the
X-ray Background are not resolved, the sources are flux weighted in
their formalism.  In our case, weighting by the flux of each radio
source can make the measurements very noisy.  Instead, we derive here
the number weighted case (cf. Piran \& Singh 1996 for a similar
formalism for Gamma-ray bursts).  We begin by writing the harmonics as
\begin{equation}
a_{l}^{m} = \int dV_c  \;
  [1 + b_R \delta_m({\bf r})] \;
\Phi(r) \;\ {Y_{l}^{m^*}}({\bf \hat r}),
\end{equation}
and the {\it fluctuations} over the mean in the distribution (for $l > 0$) as
\begin{equation}
\label{e7}
a_{l}^{m}= \int d\omega \; dr \; r^2  \;\delta_R({\bf r}) \;  \Phi(r) \;
\ {Y_{l}^{m^*}}({\bf \hat r}),
\end{equation}
where the comoving volume element $dV_c$
has been split into its radial and angular components, and
$\delta_{\bf R}$ is the fluctuation in the number of radio galaxies with
respect to the mean. 
For simplicity (reflecting our ignorance about 
the evolution of radio sources) we have assumed linear epoch-independent 
biasing factor between the radio
sources and the mass fluctuations
\begin{equation}
\delta_R({\bf r}) = b_R \; \delta_m({\bf r}).
\end{equation}

Now we make use of the Fourier relation between the spatial and wavenumber
density fluctuations. By substituting the Rayleigh expression,
\begin{equation}
e^{i {\bf k}\cdot {\bf r}} = 4 \pi \sum_{lm} i^l j_l(kr)
{Y_{l}^{m^*}} ({\bf \hat r})
{Y_{l}^{m}} ({\bf \hat k}),
\end{equation}
into eqn.~\ref{e5}, it becomes
\begin{equation}
\delta ({\bf r}) =
{1 \over {2 \pi^2 } } \sum_{lm} (i^l)^*
Y_{l}^{m} ({\bf \hat r})
 \int d^3k \; \delta_{\bf k}(z) {Y_{l}^{m^*}} ({\bf \hat k}) \; j_l(kr).
\end{equation}

By substituting this expression for the density fluctuation term in
eqn.~\ref{e7}, we get
\begin{eqnarray}
a_{l}^{m} &=& {b_R \over {2 {\pi}^2 }}  \int \int d \omega\;  
d r \; {r}^2 \; \Phi(r)\;
\sum_{l',m'} (i^{l'})^* Y_{l}^{m}( \hat{\bf r}) {Y_{l'}^{m'^*}} ( \hat{\bf r}) \nonumber \\
&&  \times\int d^3 k \; \delta_{\bf{k}}(z) Y_{l'}^{m'*}( \hat {\bf k}) \; j_l'(kr). 
\end{eqnarray}
By the orthogonality relation, $\int d \omega Y_{l}^{m} ({\bf \hat r})
Y_{l'}^{m'^*} ({\bf \hat r}) = \delta_{ll'}^{mm'}$ 
(where $\delta_{ll'}^{mm'}$ is the Kronecker  delta function), 
the $\sum$ over $l$ and
$m$ drops out of the above expression.

% OL changed notation from $q$ to $\mu$, so notation 
% agrees with LPT97
Mass density fluctuations evolve with time, growing as the scale factor
\begin{equation}
\delta_{\bf{k}}(z) = {\delta_{\bf{k}}(0) \:(1+z)}^{-\mu} .
\end{equation}
In linear theory $\mu={1}$ for an Einstein-de Sitter universe, 
and from here on we adopt this value.  

We take the mean square of the $a_{l}^{m}$'s and then the ensemble average
of the result. We use the definition below to rewrite the density fluctuation term $\langle \delta_{\bf k} \; \delta_{\bf k'}^* \rangle $ as:
\begin{equation} 
\langle \delta_{\bf k} \; \delta_{\bf k'}^* \rangle \; =  
(2 \pi)^3  P(k) \; \delta^{(3)}({\bf k} - {\bf k'} ), 
\end{equation}
where $P(k)$ is the power spectrum of fluctuations in mass  
and $\delta^{(3)}$
is the 3-dimensional Dirac delta function. 

After grouping some of the terms into the window function, $\Psi_l(k)$, we
have
\begin{equation}
\label{e2}
\langle |a_{l}^{m}|^2 \rangle = { 2 \over \pi} b_R^2\; \int dk \; k^2 P(k)
|\Psi_l(k)|^2.
\end{equation}

The expression we evaluate now gives the ensemble average (in the rms
sense) harmonic coefficients. This is the prediction for the harmonic
coefficients we would get if we averaged the $a_{l}^{m}$'s measured by all
possible observers (under the same observational conditions we have) in the
universe.  Of course, 
the radio data  represent only one realization, 
but for each harmonic $l$, we have $(2 l +1)$ independent measurements. 

To model the fluctuations we parametrize $P(k)$
for a family of Cold  Dark Matter (CDM) models.
In this case the shape of $P(k)$ is determined by 
the product of the density parameter and the Hubble constant
$\Gamma \equiv \Omega h$.
This parameter is related to  the size of the horizon 
when the densities of matter and radiation where equal.
For standard CDM $\Gamma=0.5$, while for low density CDM model
which fits empirically clustering of APM and IRAS galaxies, $\Gamma=0.2$. 
A smaller $\Gamma$ gives more power on large scales.
The normalization of the mass power spectrum is commonly given by 
$\sigma_{8,M}$, the rms fluctuations in spheres of $8 \Mpc$.

%The resulting model harmonics, for flux cutoffs of 50, 70 and 100 mJy, are shown in Figure~\ref{harm1a}.

%\begin{figure}
%\begin{center}
%\parbox{8cm}{\psfig{figure=models_c.ps,width=8cm,clip=}}
%\end{center}
%\caption{\label{harm1a} Predicted spherical harmonic coefficients, ${\langle |a_{l}^{m}|^2
% \rangle}_{mod}$, for 50, 70 and 100mJy cutoffs, for full sky coverage, and LDCDM model
%parameters.}
%\end{figure}

\subsubsection{The Window Function, $\Psi_{l}(k)$}

We have grouped the distance and redshift terms into $\Psi_l(k)$
\begin{equation}
\label{winfunc}
\Psi_l(k) = \int dr \; {r}^2 \Phi(r) j_l(kr) {(1+z)}^{-\mu} .
\end{equation}
Here $r$ is the comoving distance, which is related to redshift
by
$r= {{2c}\over {H_0}} [1 -(1+z)^{-1/2}]$.
We do not know the selection function, $\Phi(r)$, explicitly and have to 
express in terms of   
the observed redshift 
distribution.
Radio luminosity functions (Dunlop \& Peacock 1990) can be used to
calculate the number of radio sources we expect to detect as a function of
redshift, $z$, given the frequency at which we are observing and a flux
cutoff.  The resulting redshift distribution, $N(z)$, gives the number of radio
galaxies observed per steradian at a redshift, $z$.

$N(z)$ was derived 
from the Dunlop \& Peacock (1990)
radio luminosity functions for a number of flux cutoffs, as explained in Loan et al.(1997).
The luminosity range of radio galaxies is so large that the resulting
redshift range probed is approximately the same for a range of 
flux cutoffs.

From the definitions of $\Phi(r)$ and $N(z)$, we know that 
\begin{equation}
\label{b}
\int dr \;{r}^2 \Phi(r) = {\cal N} = \int dz \; B N(z) ,
\end{equation}
where ${\cal N}$ is the number of galaxies per steradian in our survey and
B is a normalisation constant. Since the above equality holds for all
${r}$, we can rewrite ${\Psi_l(k)}$ in terms of $N(z)$.

\begin{equation}
\label{psi}
\Psi_l(k) = 2BQ \int dx \; {N(z(x)) {j_l(x)} {(1-Qx)}^{-1}} ,
\end{equation}
where we have set $x=kr$ and $Q={H_o}/2kc$.

\begin{figure}
\begin{center}
\parbox{8cm}{\psfig{figure=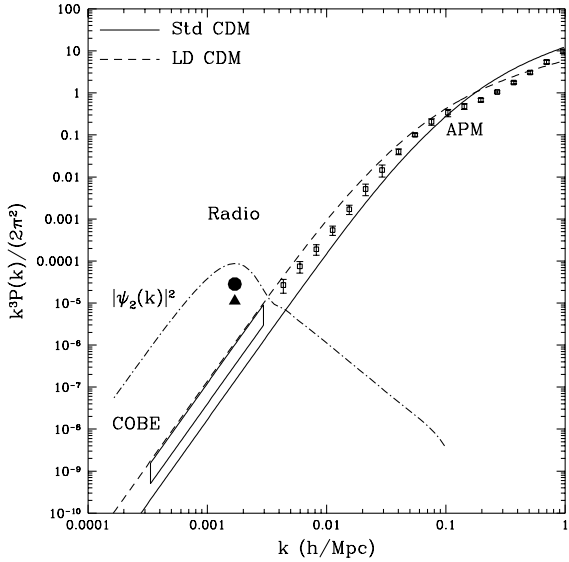,width=8cm,clip=}}
\end{center}
\caption{\label{pk} Radio surveys as probes of scales between those 
accessible to galaxy surveys (e.g. APM) and the CMB (e.g. COBE).
The solid and dashed lines represent 
$k^3 P(k)/(2 \pi^2) \sim ({{\delta \rho} \over \rho})^2$
for standard CDM ($\Gamma=0.5$) 
and low density CDM ($\Gamma=0.2$) models, 
both normalized with $\sigma_{8,M}=1$ at $k \sim 0.15$.
The open squares at large $k$'s (small scales) 
are estimates of the power-spectrum from 3D inversion
of the angular APM galaxy catalogue
(Baugh \& Efstathiou 1993, 1994).
The elongated 'box' at small $k$'s represents the COBE 4-yr CMB measurements 
(E. Gawiser, private communication). 
It corresponds to 
a quadrupole Q=18.0 $\mu K$ for a Harrison-Zeldovich mass 
power-spectrum, via the Sachs-Wolfe effect, or $\sigma_{8,M} = 1.4$ 
for a standard CDM model.
The dashed-dotted line is the window-function $|\psi_2(k)|^2$
for the quadrupole (eqn.~\ref{winfunc}) of radio sources brighter than 
70 mJy and an Einstein-de Sitter universe.
It illustrates the scales probes by the spherical harmonic 
analysis of radio surveys (eqn.~\ref{e2}). 
It peaks at $k_*^{-1} \sim 600 h^{-1}$ Mpc,  indicating the 
capability of radio surveys to probe intermediate scales.
The vertical scaling of the window function is arbitrary.
The solid triangle and circle are estimates of the power-spectrum
at $k_*$,  assuming the shape of standard CDM and LDCDM power-spectra
respectively.
They are based on the measured quadrupole from the 
combined 70 mJy ${87GB-PMN}_{match}$ sample, which correspond to 
$\sigma_{8,R} = b_R \sigma_{8,M} \sim 9 $ and 5 
for standard CDM and LDCDM respectively.
Given the problems of catalogue matching and shot-noise, 
these points should be interpreted at best as `representative 
upper limits', not as detections.  
}
%\label{pk}
\end{figure}

It is very insightful to plot the window function on top 
of the curves for 
$k^3 P(k)/(2 \pi^2) \sim ({{\delta \rho} \over \rho})^2$,
as it shows what scales of $k$ are probed by particular statistic 
applied to a specific catalogue.
The window function for the quadrupole 
$|\psi_2(k)|^2$ (eqn.~\ref{winfunc}) for 70 mJy survey, 
is shown by the dashed-dotted line in 
Figure~\ref{pk}, together with the power-spectra 
for standard CDM (with shape parameter $\Gamma=0.5$) 
and  low density CDM ($\Gamma=0.2$) models.
In both cases we assume normalization $\sigma_{8,M}=1$ and
$b_R=1$ (no biasing).
The figure illustrates that 
radio surveys can serve as  probes of scales 
$k_*^{-1} \sim 600 h^{-1}$ Mpc, 
between those accessible to 
galaxy surveys (e.g. APM) and the CMB (e.g. COBE).
Other variants of this plot which include other probes
of the power-spectrum (e.g. local redshift surveys and the X-ray Background) 
are given elsewhere (e.g.  Baugh \& Efstathiou 1993; Lahav et al.  1997; 
Wu, Lahav \& Rees, in preparation).

\subsubsection{Shot Noise}

Finally, we need to consider the discreteness of the galaxies, which
contributes a shot noise term to the data harmonics. To account for this, we add a shot noise
estimate to the predicted model harmonics.  Each measured $a_{l}^{m}$ will be
subject to the same shot noise contamination
\begin{equation}
{\langle |a_{l}^{m}|^2 \rangle}_{SN}  = {1 \over 4 \pi} \sum_{sources} 1 = {\cal 
N}, 
\end{equation}   
so
\begin{equation}
\label{almtot}
{{\langle{|a_{l}^{m}|}^2}\rangle}_{tot} =
{{\langle{|a_{l}^{m}|}^2}\rangle}_{mod} +
{{\langle{|a_{l}^{m}|}^2}\rangle}_{SN} .
\end{equation}
An added advantage of number weighted harmonics (over flux weighted) comes
in here in the shot noise term. For the number weighted case the shot noise
is finite, whereas for the flux weighted case it diverges 
if bright sources are not removed.
(cf. Lahav et al. 1997).

\subsubsection{Incomplete Sky Coverage}

The radio surveys considered  here
only cover about $70\%$ of the sky. The remaining $30\%$ is
filled in approximately uniformly (according to eqn.~\ref{e8}) to the mean density of the
surveyed sky. We do this to avoid the crosstalk between the harmonic
coefficients, $a_{l}^{m}$, that can result from an incomplete sky. As we mention in Sections 5 and 6, we still measure significantly larger than predicted harmonics amplitudes in the data.   
However, the SHA formalism presented in Section 3 is for full sky
coverage. It is possible that despite our treatment of the unsurveyed regions, we are still seeing an effect of incomplete sky coverage.  

To make sure that we are comparing the data and model harmonics on equal footing, we want to deal with the
unsurveyed sky more rigorously. This can be done in two ways.  
One is to solve the inversion problem (given the known mask, i.e. unsurveyed region geometry) and 
to retrieve the data harmonics for a complete sky.  The second approach is
to adapt the model harmonics to additionally incorporate the effect of
incomplete sky coverage and predict the rms harmonics for all observers
with the same incomplete sky we have. We use the second method to examine the  amplitude of the effect of incomplete sky coverage.

The equation for the predicted harmonics for an incompletely surveyed sky
is then
\begin{equation}
\label{tensor}
{\langle {|c_{l}^{m}|}^2 \rangle}_{tot}= \sum_{l'} \sum_{m'} {|W_{ll'}^{mm'}|}^2
\; {\langle {|a_{l'}^{m'}|}^2 \rangle}_{tot} ,
\end{equation}
where the $W_{ll'}^{mm'}$ tensor models the unsurveyed regions (see Peebles
1973, Scharf et al 1992), and the $ {\langle {|a_{l'}^{m'}|}^2 \rangle
  }_{tot}$'s are the predicted harmonics for a completely surveyed sky
(equation ~\ref{almtot}).
This can also be written as
$
{{\langle{|c_{l}^{m}|}^2}\rangle}_{tot} =
{{\langle{|c_{l}^{m}|}^2}\rangle}_{mod} +
{{\langle{|c_{l}^{m}|}^2}\rangle}_{SN},
$  
where the masked shot-noise term is
$ {\langle{|c_{l}^{m}|}^2 \rangle}_{SN} = (\Omega_{obs}/4 \pi) {\cal N} $. 

\begin{figure}
\centerline{ \psfig{figure=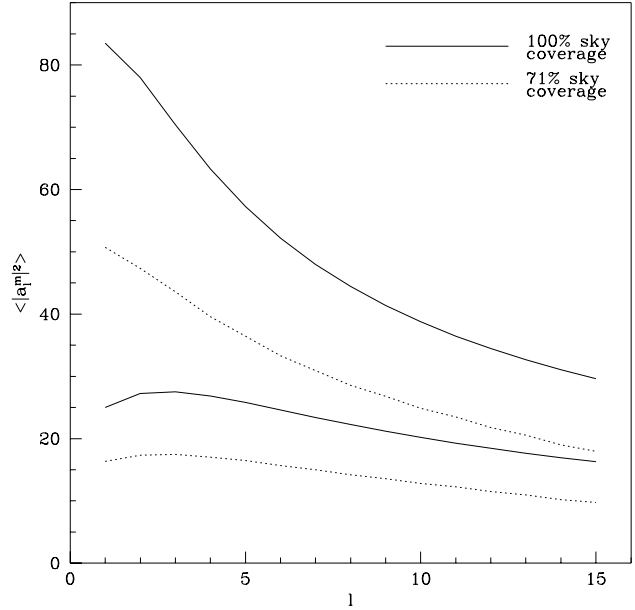,width=3.5in} }
\caption{\label{harm2} Predicted ensemble average harmonics, using N(z) for a 70 mJy
cutoff for LDCDM (higher curve) and CDM (lower curve) power spectra. Solid
lines correspond to ${\langle |a_{l}^{m}|^2 | \rangle}_{mod}$ for full sky,
and dotted lines correspond to ${\langle |c_{l}^{m}|^2 | \rangle}_{mod}$
for $70 \% $ sky coverage, for an ideal 4.85 GHz survey.}
%\label{harm2}
\end{figure}

Details of the calculation of the $W_{ll'}^{mm'}$ tensor are given in
Scharf et al (1992), and Zare (1987). The result of this correction is a
lowering of the predicted ensemble average amplitudes, not an increase, as might have explained the large observed harmonic amplitudes. The corrected amplitudes $ {\langle{|c_{l}^{m}|}^2 \rangle}_{mod}$ in the absence of shot-noise for a 70 mJy flux cutoff, are shown in
Figure~\ref{harm2}, compared to the prediction for full sky coverage
$ {\langle {|a_{l}^{m}|}^2 \rangle}_{mod}$.
This shows the trend in predicted harmonic amplitudes between the CDM and LDCDM power spectra and between full vs. partial sky coverage for a single flux cutoff. There is also a trend across chosen flux cutoff. With respect to the amplitudes shown for a 70 mJy cutoff, those for 50 mJy and 100 mJy are, repectively, $\approx .5 \times$ and $\approx 2 \times$ as high.

\section{The Predicted Dipole}
\label{dipole}

The $l=1$ harmonic term has additional significance in the study of large
scale structure. The well known CMB dipole anisotropy is thought to arise
from our peculiar velocity relative to the CMB. This motion
is attributed to the gravitational pull of 
anisotropically distributed mass overdensities around us.
In linear theory (Peebles 1980) the
peculiar velocity vector, {\bf{v}}, is proportional to the gravitational
acceleration vector, {\bf{g}},
\begin{equation}
{\bf{v}} \propto {{\Omega^{0.6}} \over b} {\bf{g}},
\end{equation}
where $\Omega$ is the density parameter at the present epoch and $b$ is the
bias parameter (assuming linear biasing).

%\subsection{Theoretical Dipoles}

Galaxy surveys do not
represent the total mass directly,
although the light emitted by galaxies
at certain frequencies is likely 
to be indicative of their mass and thus related to the underlying mass distribution.
The dipole can be calculated
as a flux weighted sum over each galaxy's position (see eqn.~\ref{dipgen} for a more general form).
If the  galaxy's light is proportional to its mass
this  can  represent the 
gravity vector as both the flux and gravitational force fall off with distance as $1/r^2$.

In practice, however, all available galaxy catalogues are flux limited, and
detect only a fraction of the galaxies in any volume surveyed. This means
that the calculation will give an estimate of the dipole which
is produced by only those galaxies detected out to a limited effective
distance.
Moreover, in this study, we have fluxes at radio wavelengths and the radio emission from galaxies may not be a good indicator of galaxy mass.  
We therefore 
prefer to calculate 
a number-weighted dipole.
As a result, we cannot expect the number-weighted  dipole 
to be aligned with the mass gravity vector and with 
the CMB Local Group velocity dipole.

We now consider the various effects that may contribute to the dipole
amplitude measured for a population of radio sources.  We parameterise the
dipole as
\begin{equation}
{dN \over {d \Omega}}(\theta) = {\cal{N}} + A \cos \theta,
\end{equation}
where the left side of the equation is the mean density of objects
measured, per unit solid angle, as a function of direction, $\theta$, on
the sky.  The angle $\theta$ is measured from the direction of peak dipole
effect. ${\cal{N}}= {N \over 4\pi}$ is the number of galaxies per steradian
in the combined 87GB-PMN catalogue for a given flux cutoff. $A$ is the dipole
amplitude we expect to measure due to the effects discussed below. The
different notation for the dipoles is related as follows:
\begin{equation}
A= {3 \over {4 \pi}} D = {3 \over \sqrt{4 \pi}} \sqrt{\langle
{|{a}_{1}^{m}|}^2 \rangle},
\end{equation}
where $D = |\sum_i {{\bf \hat r_i} }| $ is the amplitude of the number-weighted 
dipole calculated and listed in
Tables ~\ref{initdip} and ~\ref{newdip}, and ${\langle {|{a}_{1}^{m}|}^2
  \rangle}$ is the dipole term of the harmonic prediction of our models
 (Section~\ref{sha}).

As described below,
the velocity dipole is a separate effect from the large scale
structure dipole, although related to it. The large scale structure induces
our motion, which in turn creates the blueshifted excess of galaxies in the
direction of motion. The measured dipole will be a combination of these two
effects, plus the shot noise contribution.

\subsection{Velocity term} 

For the sake of the following calculation, the population of radio galaxies
is assumed to be isotropic in its rest frame.  Ellis and Baldwin (1984)
predicted the dipole amplitude we should measure in a flux limited radio
galaxy survey, due to our motion with respect to the rest frame of the radio galaxy distribution.  Consider a population of radio galaxies, whose radio continua can be characterized by the power law $S \propto {\nu}^{- \alpha}, \alpha$ positive. 
In the direction of the sun's motion, galaxies are
blueshifted, so the observed 4.85 GHz band actually samples a lower
frequency band in the rest-frame of the galaxy, and hence a higher
flux. 
%In the direction of motion, galaxies that were too faint (in
%their rest frame) to be detected at our chosen flux limit, are ``boosted''
%in flux so that we can see them. In the opposite direction we lose sight of
%galaxies that would be just above the flux limit if we were not moving. 
The overall effect is that we see more galaxies in the direction of motion and
less in the opposite direction than we would if we were not moving with
respect to the galaxies' rest frame. This deviation
from isotropic number counts introduces a dipole pattern into the galaxy
counts. This is similar to the 'Compton-Getting' effect flux dipole 
due to the motion of the observer relative to a sea of radiation 
(e.g. the CMB or the X-ray Background).

This velocity dipole effect was predicted  by Ellis \& Baldwin (1984) as
\begin{equation}
A_{vel} = [ 2 + x ( 1 + \alpha ) ] {v \over c} \; {\cal{N}} ,
\end{equation}
where $x$ is the slope of the integral source count per unit solid angle,
above a given limiting flux, $S_{lim}$,
\begin{equation}
{dN \over d{\Omega}}(S > S_{lim}) \propto {S}^{-x}.
\end{equation} 
As mentioned above, $\alpha$ is the slope of the power law spectra of the
surveyed objects, and $v$ is our velocity with respect to the frame
in which the counts are isotropic.  Note, that the above estimate is for a
number weighted dipole. We take $x =1$ ( $=1.5$ in a Euclidean universe), and $
\alpha = 0.75$ (mean spectral index of radio galaxy spectra)
). The velocity
inferred from the CMB dipole is $v_{sun-CMB} \sim 370 {km \over s}$. The
velocity dipole effect we should measure is
\begin{equation}
\label{vel}
A_{vel} = 4.625 \times {10}^{-3}{\cal{N}}.
\end{equation}

As an example of the magnitude of the effect, in
the combined catalogue used below we have ${\cal {N}} = 2878$
for $S_{lim} = 70 mJy$. In this case:
\begin{equation}
A_{vel}= 13.3
\end{equation}

In principle, we 
can subtract the velocity dipole  based on the solar motion 
with respect to the CMB, and then consider the residual dipole 
as being purely due to large scale structure.
It should be noted  that the $A_{vel}$, unlike $A_{LSS}$ and $A_{SN}$
discussed below, is
{\it not}  an rms quantitity with zero mean, but in fact a correctable
value.  It is thus quite different in character from the
other two terms.
Eq. (23) only holds for a single population of radio sources 
characterized  by spectral index $\alpha$ and counts index $x$.
In reality, the 4.85 Ghz sample is composed of different populations, 
and therefore the expression for $A_{vel}$ should be a superposition 
due to the different populations. At present the distribution functions for
$\alpha$ and $x$ at 4.85 GHz are poorly known, so it is difficult to 
evaluate the exact prediction for $A_{vel}$.

At this point, it is important to emphasize the distinction between the two reference frames with respect to which the CMB dipole amplitude and direction are defined. One is the heliocentric frame: $v_{sun-CMB} \sim 370 {km/s}$
towards Galactic coordinates $l=264^\circ; b=48^\circ$. As the measurements are done in the heliocentric frame, $v_{sun-CMB}$ will determine the amplitude of $A_{vel}$.  
The other relevant reference frame is the Local Group frame: $v_{LG-CMB} \sim 600 km/s$ toward
 $l=268^\circ;b =27^\circ$. 
Since the radio galaxies in the catalogues are outside the Local Group, this pertains to the LSS dipole prediction in our discussion, 
but as explained above there is no simple relation between a number-weighted 
dipole and the CMB Local Group velocity dipole.

\subsection{Large Scale Structure term}

This is the $l=1$ term, from the predicted spherical harmonics in
Section~\ref{sha}.
 
As an example of the magnitude of the effect we 
have assumed a power
spectrum which fits the observed local clustering of optical and IRAS
galaxies, parameterised as an unbiased, low density CDM (hereafter LDCDM) 
power-spectrum with 
shape parameter $\Gamma=0.2$.
We also assume an Einstein-de Sitter universe ($\Omega =1$)
and hence that 
fluctuations grow as $(1+z)^{-1}$. 
We adopt a redshift distribution, $N(z)$, corresponding to 
a flux cutoff of 70mJy.
For these parameters the rms prediction, eqn.~\ref{e2},
gives ${\langle {|{a}_{1}^{m}|}^2 \rangle} = 83.52$, 
which translates into 
\begin{equation}
A_{LSS}= 7.7.
\end{equation}

\subsection{Shot Noise term}

As mentioned earlier, a shot noise term comes in because we are using discrete
galaxies as tracers of a continuous quantity (fluctuations in the mass
density). We calculated the Poisson shot noise contribution to be
\begin{equation}
{\langle |a_{l}^{m}|^2 \rangle}_{SN} = {1 \over 4 \pi} \sum_{sources} 1 =
{\cal N} ,
\end{equation}   
so
\begin{equation}
\label{sn}
A_{SN} = {3 \over {\sqrt{4 \pi}}} \sqrt{\cal{N}}.
\end{equation}

In the case of our combined catalogue
for a 70mJy flux cutoff, ${\cal{N}}= 2878$ and we expect: 
\begin{equation}
A_{SN} = 45.4.
\end{equation}

Unfortunately, the shot noise turns out  be the dominant component of any
dipole we measure in the 87GB-PMN survey. 
It is simple to estimate the number of objects we would need to detect the
velocity and large scale structure dipoles at the same level as shot noise.
For the velocity dipole, we equate eqns.~\ref{vel} and ~\ref{sn} and find
that we need
\begin{equation}
{\cal{N}}
\approx \; {3.35 \times {10}^{4} \;galaxies \; per \; str} ,
\end{equation}
which translates into $N \sim {4 \times {10}^{5}}$ galaxies over the sky. The large scale
structure dipole, $A_{LSS}$ is proportional to $\cal{N}$ (see
eqns.~\ref{e2},~\ref{b},~\ref{psi}), so by equating eqn.~\ref{sn} and the
expression for the large scale structure dipole 
we find we need
\begin{equation}
{\cal{N}} \approx \; {1.06 \times {10}^{5} \;galaxies \; per\; str} ,
\end{equation} 
(or $N \sim 1.3 \times {10}^{6}$ galaxies total) to be able to detect a
large scale dipole (for the LDCDM power spectrum, the shape of $N(z)$ 
and the other parameters mentioned
above).

The  predicted velocity, large scale structure, and
shot-noise dipole effects (for the example parameters 
mentioned above) are summarized in Table ~\ref{thdip}.  
Our a priori calculation shows that we are unlikely to detect
a dipole (and higher harmonics, see below) 
in the 87GB-PMN catalogues due to the 
high level of shot-noise expected.

Nevertheless, we have attempted to calculate the dipole and higher harmonics
from the catalogues for several reasons: 
(i) The  detection of the 
angular correlation function (Loan et al. 1997) 
and the Supergalactic Plane (Section 7) 
in ${87GB}_{raw}$ and ${PMN}_{raw}$ suggests the existence of large scale structure.
(ii) It may well be that our formalism of Gaussian random fields
does not fully characterize real structure.
(iii) Comparison of the predictions to observations can serve as
a consistency check of the validity of the ${87GB}_{raw}$ and ${PMN}_{raw}$ flux
calibration.
(iv) The application to existing catalogues can serve 
 as a pilot study for future deeper surveys.

As described below, we find that the observed dipole is much higher
than expected. We attribute this discrepancy to the flux matching 
of the catalogues.
However, it is interesting that after applying an ad-hoc flux matching 
procedure we still detect a dipole (see Table ~\ref{thdip}) 
larger than predicted. Flux cutoffs of 50mJy and 100mJy result in qualitatively similar results - the detected dipole is still higher than the predicted shot noise, large scale structure and velocity dipoles. The magnitude of this surplus of the detected dipole with respect to the predicted dipoles increases with decreasing flux limit. We discuss the implications for this detection in the next few
sections.

\begin{table}
\caption{\label{thdip} The amplitudes of various dipole effects, compared to the 
number weighted dipole we detect (after the $\cal N$ matching procedure) in the ${87GB-PMN}_{match}$ catalogues, for a 70mJy flux 
cutoff.}

\centerline{
\begin{tabular}{|l|l|}
\hline
\multicolumn{1}{|l|}{Dipole} & \multicolumn{1}{|c|}{A (amplitude)}\\
\hline\hline
Velocity & 13.3 \\
Large Scale Structure & 7.7 \\
Shot Noise & 45.4  \\
\hline
Detected & 73.3 \\
\hline
\end{tabular}
}
\end{table}

\section{Dipole calculation using the combined ${87GB-PMN}_{match}$ catalogue }

Despite the numerous uncertainties in relating the radiation emitted by
galaxies to the underlying mass distribution, angular dipole calculations
for a variety of galaxy catalogues (e.g. Meiksin \& Davis 1986; Yahil,
Walker \& Rowan-Robinson 1986; Villumsen \& Strauss 1987; Lahav 1987;
Harmon, Lahav \& Meurs 1987; Lahav, Rowan-Robinson \& Lynden-Bell 1988;
Plionis 1988; Lynden-Bell, Lahav \& Burstein 1989; Kaiser \& Lahav 1989;
Scharf et al. 1992), have found amazingly good alignment
($10^{\circ}-30^{\circ}$) to the CMB dipole in the Local Group 
frame. Similar alignment has been
detected using IRAS redshift surveys (Rowan-Robinson et al. 1991; Strauss et
al. 1992; Webster, Lahav \& Fisher 1997) 
and optical redshift surveys (e.g. Hudson 1993).

The detection of a dipole closely aligned with the CMB dipole, in a number
of surveys at various wavelengths, suggests that it is worth calculating
the dipole for the radio galaxies in the 87GB-PMN survey. We only have angular
coordinates and fluxes.  The 87GB-PMN surveys have a median
redshift of $z \sim 1$.  This is important for the question of convergence of
the measured galaxy dipole to the CMB dipole. Above some large enough
distance from us, we expect the distribution of galaxies to be homogeneous,
and not contribute to the CMB dipole.  The 87GB-PMN catalogues are deep
enough to probe the convergence of the dipole far beyond the limit set by
IRAS and optical surveys. 

On the other hand, we do not know how well radio
radiation from a galaxy indicates its mass, or how radio galaxies are
biased with respect to the underlying mass distribution.  In addition, the
sparse sampling of the 87GB-PMN surveys may make it difficult to even
detect a dipole signal. If we assume an Einstein-de Sitter cosmology, for a
70mJy cutoff there are only $\sim 500$ galaxies out to $100 \Mpc$ and $\sim
1500$ objects out to $z \sim 0.1$ in the combined radio catalogues.

After filling in the unsurveyed regions of the sky, we calculated the radio galaxy dipole for a range of flux cutoffs in the
survey.  The general form of the dipole calculation (for a full sky
distribution of N galaxies) is
\begin{equation}
\label{dipgen}
{\bf{D}}=\sum_{i=1}^{N} w_{i} \hat{r_{i}} .
\end{equation}

For a flux weighted dipole, $\it {w_{i}} = \it {f_{i}}$, the flux of galaxy
$\it {i}$.  We use $\it {w_{i}} = 1$, and calculate the number weighted
dipole because, although the flux weighted dipole is the optimal
calculation to recover the acceleration vector, it is not always reliable.
As mentioned above, only some wavelengths of a galaxy's radiation are good
indicators of its mass. If there is no correlation between $L_{gal}$ at a given wavelength and
$M_{gal}$, the flux weighted dipole will not give an estimate
of the gravitational pull we are experiencing. In such a case, it is better
to use a strictly number weighted dipole. Note that this no longer gives an
estimate of ${\bf g}$ because it does not incorporate the $1/r^2$ fall off in flux to each
galaxy (as with $w_{i} =f_{i}$), and as such, is no longer directly comparable to the LG-CMB dipole. The results of our initial calculation are shown in Table ~\ref{initdip}. 
%The %dipoles point $\sim 50\degr$ from the LG-CMB dipole. 
As mentioned, no alignment was expected. The measured amplitudes, however, are significantly larger than the number weighted prediction from Section 3.

\begin{table}
\caption{\label{initdip} Initial dipole calculation for ${87GB-PMN}_{raw}$. $S_{min}$ is the flux cutoff in mJy, $\cal{N}$ is the number of galaxies per steradian, $D$ is the
amplitude of the dipole, and $l$ and $b$ give the Galactic coordinate 
direction in which the dipole is pointing.}
\centerline{
\begin{tabular}{|c|c|l|l|r|}
\hline
\multicolumn{5}{|l|}{Cumulative dipole (${w_i}=1$)}\\
\hline\hline
\multicolumn{1}{|c|}{$S_{min}$} & \multicolumn{1}{c|}{$\cal{N}$} & \multicolumn{1}{c|}{D} & \multicolumn{1}{c|}{l} & \multicolumn{1}{c|}{b}\\
\hline
50 & 4671 & 1342.3 & $299^{\circ}$ &$ -21^{\circ}$ \\
70 & 2985 & 709.2 & $297^{\circ}$ & $-15^{\circ}$ \\
100 & 1831 & 468.4 & $300^{\circ}$ &$ -17^{\circ}$ \\
\hline
\end{tabular}
}
\end{table}

Note that the above dipole calculations have been done
under the assumption that the ${87GB}_{raw}$ and ${PMN}_{raw}$ catalogues were produced using
similar reduction criteria, and that for each flux cutoff ${{\cal{N}}_{PMN-raw}
= {\cal{N}}_{87GB-raw}}$ (where $\cal{N}$ is the number of galaxies per
steradian on the sky). As mentioned in Section~\ref{radio}, this is not a
valid assumption. For the combination of the ${87GB}_{raw}$ and ${PMN}_{raw}$ catalogues that
we use, we have found the regions of the sky covered by the ${PMN}_{raw}$ catalogue to
have larger $\cal{N}$ values for a given flux cutoff.

The resulting dipole in the flux matched catalogue, for a
70mJy flux cutoff was $D = 307.2$, a significant decrease from the original value. The direction also changed considerably, to $l=286^{\circ},
b={15}^{\circ}$, $21\degr$ away from the LG-CMB dipole, $38\degr$ away from the Sun-CMB dipole.

It is interesting to note that our derived dipole (Table ~\ref{newdip}) 
is closer to to the Local Group CMB dipole than to the heliocentric 
CMB dipole. 
%$(l=264^\circ; b = 48^\circ)$. 
%E.g. the angle between our dipole and the heliocentric CMB dipole
%is $18^\circ, 70^\circ$ and $38^\circ$ for flux limits of 
%50,60 and 70 mJy respectively (compared with   
%$15^\circ, 51^\circ$ and $21^\circ$ relative to the Local Group 
%CMB dipole). 
As previously mentioned, however, 
not much can be inferred from the alignment between our number-weighted dipole 
and the LG-CMB dipole.
At most, this suggests that the velocity dipole does not dominate the measured dipole.

\begin{figure}
\begin{center}
\parbox{8cm}{\psfig{figure=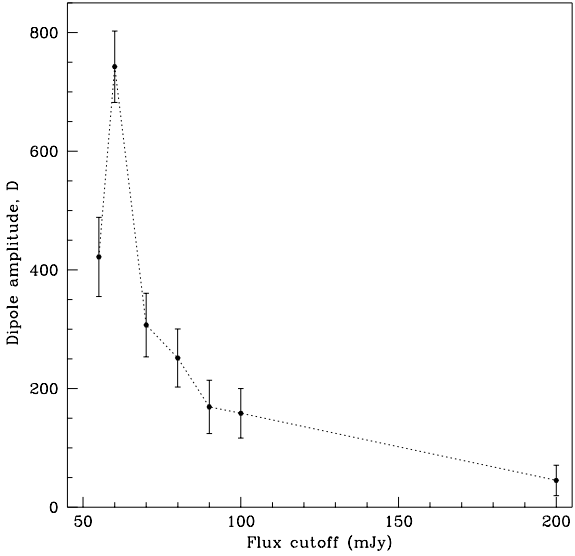,height=8cm,clip=}}
\end{center}
\caption{\label{figdip1} After $\cal N$ matching : Detected dipole amplitude, D,
for flux cutoffs from 50 mJy to 200 mJy.}
\end{figure}

\begin{figure}
\begin{center}
\parbox{8cm}{\psfig{figure=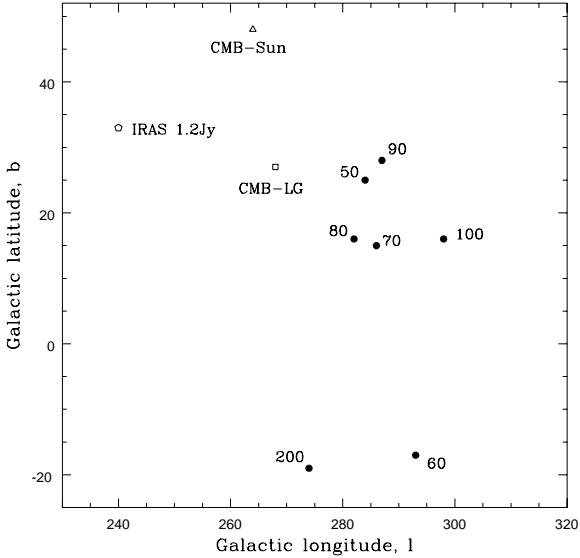,height=8cm,clip=}}
\end{center}
\caption{\label{galdip} After $\cal N$ matching : Detected dipole direction, 
for flux cutoffs from 50 mJy to 200 mJy (filled circles), the IRAS 1.2 Jy dipole (open pentagon), LG-CMB dipole (open square), and Sun-CMB dipole (open triangle).}
\end{figure}

The decrease in the measured dipole amplitude  was encouraging enough to
recalculate the dipole for other flux cutoffs.  The results for a range of flux cutoffs are shown in Table ~\ref{newdip}
and Figures ~\ref{figdip1} and ~\ref{galdip}.  The dipole amplitudes are smaller by a factor of
$2-3$, and show much closer alignment to the CMB dipole than before. For
the 60 mJy cutoff, the dipole amplitude is still quite high. We
attribute this to the poor flux matching for this flux cutoff which left a number density mismatch
comparable to some of the original ${\cal{N}}_{PMN-raw},{\cal{N}}_{87GB-raw}$
percentage differences at other flux cutoffs.

We see a significant reduction in the amplitude and change in the direction
of the dipole, after applying fairly naive (and seemingly small) changes to
the catalogues. It is possible that 
the catalogue geometry (which we have tried to
correct for by approximately uniform filling in of the unsurveyed regions)
is causing a residual dipole signal.  

\begin{table}
\caption{\label{newdip} Recalculated dipole - $S_{min}$ is the flux cutoff in mJy,
$\cal{N}$ is the average number of galaxies per steradian in the ${87GB-PMN}_{match}$ catalogues, $D$ is the amplitude of the dipole, and $l$ and $b$ give
the direction in which the dipole is pointing in Galactic coordinate. }
\centerline{
\begin{tabular}{|c|c|l|l|r|}
\hline
\multicolumn{5}{|l|}{Cumulative dipole ($w_{i}=1$)}\\
\hline\hline
\multicolumn{1}{|c|}{$S_{min}$} & \multicolumn{1}{c|}{$\cal{N}$} & \multicolumn{1}{c|}{D} & \multicolumn{1}{c|}{l} & \multicolumn{1}{c|}{b} \\
\hline
50 & 4443 & 415.3 & $284^{\circ}$ &$ 25^{\circ}$ \\
%\hline
60 & 3622 & 742.7 & $293^{\circ}$ &$ -17^{\circ}$ \\ 
%\hline
70 & 2878 & 307.2 & $286^{\circ}$ & $15^{\circ}$ \\
80 & 2392 &  251.7 & $282^{\circ}$ & $16^{\circ}$ \\
90 & 2016 & 169.2 & $287^{\circ}$ & $28^{\circ}$ \\ 
100 & 1753 &158.4 & $298^{\circ}$ & $16^{\circ}$ \\
200 & 656 & 45.3 & $274^{\circ}$ & $-19^{\circ}$ \\
\hline
\end{tabular}
}
\end{table}

\subsection{Summary and discussion of dipole calculation}

For the radio sources in the 4.85 GHz, 87GB-PMN combined catalogue, for a flux cutoff of 70 mJy, we find the following: 
\begin{enumerate}
\item{We predict the effects of our motion and of large scale structure to be
 comparable 
in amplitude to within a factor of 2.}
\item{We expect shot noise to dominate any dipole in the 87GB-PMN 
catalogues.}

\item{When we measure the dipole we find that, 
\begin{itemize}
\item{For no correction of the catalogues, $A$ is $3.5\sigma$ above the noise, but over $50\degr$ away from the LG-CMB dipole.}
\item{When we try the flux matching procedure, we measure a dipole signal of only $0.5\sigma$ above noise, but now the
dipole points $20\degr$ away from the LG-CMB dipole. }

\end{itemize}}

\end{enumerate}

We attribute most of the initially measured dipole signal to the fact that
two different reduction algorithms were used in compiling the 87GB-PMN
catalogues. The signal decreases after flux matching, but is still above
shot noise level. In addition, we now see closer alignment with the
LG-CMB dipole, that we should not detect a priori. What might produce these effects?

We hope that, after the flux matching done, the ${87GB-PMN}_{match}$ catalogue is
consistent, and not producing additional significant dipole effects,
resulting from the slight, remaining $\cal{N}$ mismatch. However, as
mentioned, there is the possibility that survey geometry is significantly influencing the dipole alignment.

An alternate explanation of the dipole detection over what we predict,
could come from the large scale structure prediction.  The large scale
structure dipole we predict is for an ensemble average of observers over a
universe (with LDCDM power spectrum). The dipole measurement of the
87GB-PMN catalogue represents only one realization, which may
greatly differ from the ensemble average. We may live in a highly
non-homogeneous part of the universe. 
There is the possibility 
that we are detecting
real large scale structure. 
As we report in Section 7 we detect a slight
increase in the mean density of sources in the ${87GB}_{raw}$ and ${PMN}_{raw}$ catalogues
within $10\degr-15\degr$ along the Super Galactic Plane. This
structure may contribute to the remaining dipole signal we measure. 

Some support for a dipole pattern in the distribution of radio galaxies
comes from an analysis by Lahav \& Shaver (1991, unpublished).
They analysed an optically-radio selected
sample (Shaver 1991) of 92 galaxies which
flux at  408 MHz larger than 1 Jy and blue magnitude brighter than
14.5, 2/3 of them with redshift
$z <0.02 $. If the dipole is so local, it is due to only few hundreds 
of sources (see our discussion of the SGP below).

%When the contribution to the dipole is weighted by the optical blue flux
%(eqn.~\ref{dipgen}) and hence 
%related to the acceleration vector,
%the resulting dipole points towards $(l=293^\circ; b=27^\circ)$,
%$22^\circ$  away from the Local Group CMB dipole.
%However, the above  calculation made no correction for the radial selection
%function and for the Zone of Avoidance.

\section{SHA application}

We now compare observations and predictions for higher harmonics.
The fluctuations in radio sources can be related to the 
fluctuations in mass via the linear bias parameter $b_R$, 
as $\sigma_{8,R}=  b_R \sigma_{8,M}$. The models we use for our predictions are for a universe
with $\Omega=1$ geometry, but we allow 2 possible power-spectra: 
 (i) standard CDM
power spectrum with $\Omega=1.0$, $h=0.5$, and normalization 
specified by the rms fluctuations in radio counts in 8 $h^{-1}$ Mpc spheres, 
$\sigma_{8,R}= 1.0$ and
 (ii) a power spectrum parameterised as an unbiased ($\sigma_{8,R} = 1.0$) 
low density CDM (LDCDM)
spectrum  shape parameter
$\Gamma=0.2$ (Bardeen et al. 1986).

We can rewrite the corrected harmonic coefficients as 
\begin{equation}
{C_{l}}^2 ={1 \over {{\cal{N}} {(\Omega_{obs}/4 \pi)} (2l+1)}}{ \sum_{m}
  |c_l^m|^2_{tot}} = 1+ 
{  \langle {|c_{l}^{m}|}^2 \rangle_{mod}  \over
{{\cal{N}} {(\Omega_{obs}/4 \pi)}}  }
\end{equation}
for $|c_l^m|^2_{tot}$ being the predicted harmonics (signal+noise,
for an incomplete sky) or
the observed harmonics. Subsequently, $C_l$ tells us directly how many
standard deviations above the shot noise a measurement or prediction is.
 The predicted harmonic 
in the absence of noise
$\langle {|c_{l}^{m}|}^2 \rangle_{mod}$ and
$C_l$ (including noise)
for a flux cutoff of 70mJy and both power spectra are listed in Table ~\ref{model}. Incomplete sky coverage has been corrected for, according to eqn.~\ref{tensor}.

\begin{table}
\caption{\label{model} The predicted harmonic coefficients for CDM and LDCDM models
for a flux limit of 70 mJy, for $l=1$ to
$l=5$. }
\centerline{
\begin{tabular}{|c|c|c|c|c|}
\hline
{ } & \multicolumn{2}{|c}{CDM} & \multicolumn{2}{|c|}{LDCDM} \\
\hline
{l} & {$\langle {|c_{l}^{m}|}^2 \rangle_{mod}$} & {$C_l$} & 
{$\langle {|c_{l}^{m}|}^2 \rangle_{mod}$} & {$C_l$} \\ \hline
   1 &   16.4 & 1.0040 &  50.7 & 1.0120 \\
   2 &   17.3 & 1.0042 & 47.3 & 1.0116 \\
   3 &   17.5 & 1.0043 & 43.6 & 1.0106 \\
   4 &   17.0 & 1.0042 &  39.6 &  1.0096 \\
   5 &   16.5 &  1.0040 & 36.4 &  1.0088 \\
\hline
\end{tabular}}
\end{table}

%\begin{figure}
%\centerline{ \psfig{figure=match_un70.ps,width=3.5in} }
%\caption{\label{fig70100} Angular power in 87GB-PMN catalogues for a 70mJy flux cutoff, before (dotted line) and after (solid line and dots) $\cal N$ matching procedure. $C_l =1$ is the shot noise level. Note that the predicted
%harmonic $C_l$'s, in columns 3 and 5 of Table ~\ref{model}, would be indistinguishable by eye, on this scale, from the shot noise line.}
%\label{fig70100}
%\end{figure}

The models predict (see Table ~\ref{model}) that due to our observing constraints,
we will not see much power (if large scale structure exists in CDM or LDCDM
form) over shot noise for any order $l$. The shot noise level is at $C_l =1$.  However, we have already measured
a larger than predicted dipole amplitude in the ${87GB-PMN}_{match}$ data, 
so we are interested in what the data
harmonics will show.

Before $\cal{N}$ matching, for the 87GB-PMN catalogue, we see a large
dipole ($l=1$) and significant signal in harmonics up to $l=10$. The ${87GB-PMN}_{match}$ data show amplitudes somewhat reduced (mainly for the lower harmonics $l < 6$, but they are still far
above the predicted amplitudes.  Even for a full sky, filled to the
same mean density as for the 70mJy flux matched case, the harmonic
coefficients oscillate considerably around the shot noise prediction.  It
is also interesting to note that the flux matched data harmonics have
higher overall amplitudes the lower the flux cutoff, following the general
trend predicted in the models.

\begin{figure}
%\label{exp}
\begin{center}
\parbox{8cm}{\psfig{figure=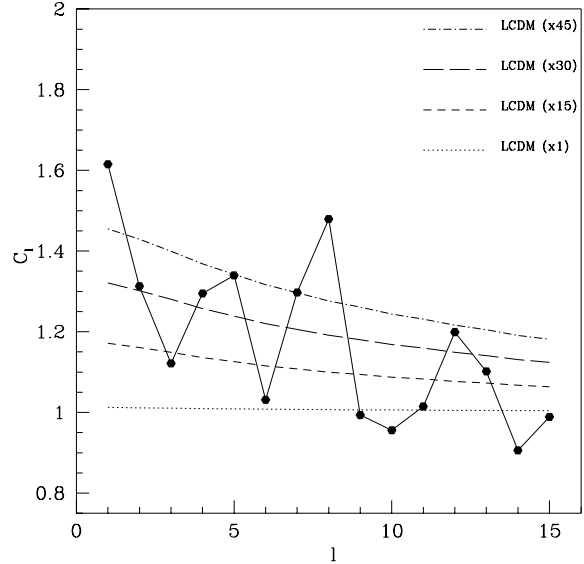,height=8cm}}
\end{center}
\caption{\label{exp} $C_l$ comparison for ${87GB-PMN}_{match}$ 70mJy data and LDCDM model predictions
(${{\langle {|c_{l}^{m}|}^2 \rangle}_{mod}}\times 1,\; \times 15, \; \times
30$ and $\times 45$) The dots connected by a solid line indicate the data
harmonics, while the various dashed and dotted lines indicate the model
harmonics. $C_l =1$ is the shot noise level. }
\end{figure}

Figure ~\ref{exp} shows the harmonics for the ${87GB-PMN}_{match}$ catalogue 
with a 70mJy
flux cutoff, compared to the predicted harmonics for the same flux cutoff,
corrected for the incomplete sky coverage. For the model prediction, we
plot the resulting $C_l$'s for ${{\langle {|c_{l}^{m}|}^2 \rangle}_{mod}}
\times 1, \times 15, \times 30$, and $\times 45$, and can see that the
galaxy harmonics are consistent with being about a factor of 15 to 30 above
the prediction (including shot noise).

The above predictions for the radio harmonics assume no biasing, 
i.e. the normalization is fixed to $\sigma_{8,R} = b_R \sigma_{8,M} = 1$. 
If we interpret the observed, $\cal N$ matched harmonics 
as being real,  we can use them to estimate the amplitude 
$\sigma_{8,R}$ for a given shape of the power-spectrum.
For example,  for the quadrupole ($l=2$) and flux limit of 
70 mJy the observation $C_2 \sim 1.3$ 
can be made to agree with the CDM and LDCDM predictions if 
$\sigma_{8,R} \sim 9$ and $\sim 5$ respectively.
(With better data one should estimate the normalisation using 
Maximum Likelihood, e.g. Scharf et al. 1992; Fisher et al. 1994.)
These crude results
 suggest a very high bias parameter $b_R$, but in qualitative agreement
with the measurement of high correlation length for radio sources.
For example, for 87GB and PMN at redshift $z=0$, 
$\xi(r) = (r/r_0)^{-1.8}$ with 
$r_0 = 18 h^{-1}$ Mpc 
(Loan et al. 1997), assuming  stable clustering.
This
can be translated (Peebles 1980) 
to $\sigma_{8,R} \approx { \sqrt {1.86 (r_0/8)^{1.8}} } \sim 2.8$.  

Figure ~\ref{pk} shows 
 (solid triangle and circle) the estimates of the power-spectrum
at $k^{-1} \sim 600 h^{-1}$ Mpc,  
assuming the shape of standard CDM and LDCDM power-spectra
respectively.
They are based on the measured quadrupole from the 
combined 70 mJy ${87GB-PMN}_{match}$ sample.
Given the problems of catalogue matching and shot-noise, 
these points should be interpreted at best as `representative 
upper limits', not as detections.

% From Andy Loan's thesis
%%%%%%%%%%%%%%%%%%%%
\section{The Supergalactic Plane (SGP)}
\label{radio-sgp-sec}
%%%%%%%%%%%%%%%%%%%%

The large-area uniformity of the 87GB-PMN surveys was adopted as the criterion for a sensible choice for
sky mask and flux-density limit.  This produced the masked catalogue used
to calculate the dipole and higher harmonics in this study, 
as well as the angular correlation function in Loan et al. (1997).  
Nevertheless, nearby
structures, such as the Supergalactic Plane (SGP), may introduce detectable
inhomogeneities into the catalogues on large angular scales.

The SGP is an over-dense region of the local
universe.  William Herschel was the first to note that `nebulae' seemed to
be more concentrated in a band across the heavens.  More recently, the
Virgo cluster was seen as the centre of a `metagalaxy' or `Local
Supercluster'.  De Vaucouleurs (1975) recognized a great circle along which
there was an increased density of galaxies in the Shapley-Ames optical
galaxy catalogue.  This great circle defines a system of Supergalactic
spherical coordinates, in which the SGP lies along the equator and the
north pole lies at Galactic coordinates ($\ell=47.37\degr,\,b=6.32\degr$).
The SGP is readily apparent as a region of greatly enhanced surface density
of galaxies in more recent large-area surveys of nearby optical galaxies
(\eg Lynden-Bell \& Lahav 1988).  Although the SGP seems to be a local
phenomenon, perhaps analogous to the sheets and filaments discovered by
galaxy surveys, it is not clear where it ends.  Studies of the SGP using
the IRAS 1.2 Jy and ORS surveys (Lahav et al., in preparation) 
show that the SGP is not a simple planar structure.

%\subsection{Moments Approach}

%\begin{equation}
%I_{ij} = \int \sigma({\bf \hat r} ) {\hat x_i} {\hat x_j} d \omega
%\end{equation}
%where $i,j=1,2,3$ label the axes.
%By substituting the surface number density $\sigma$ 
%(eqn.~\ref{eqsig}) in the above 
%equation we can express the elements $I_{ij}$ analytically 
%by the spherical harmonic coefficients of the monopole ($a_{00}$)
%and the quadrupole ($a_{2m}$), in analogy with 
%the 3-dimensional case given in 
%(Webster, Lahav \& Fisher 1997; Lahav et al., in preparation) 
%The matrix $I_{ij}$ is then diagonalised and the 
%smallest eigen-vector indicates the direction of the normal to 
%the `plane', while the eigen-values can be used to estimate the width 
%of the `plane'. 

\subsection{Histograms approach}

In the case of full sky coverage, the great circle along which 
the density of galaxies is enhanced can be found by 
calculating the 'moment of inertia' 
(Webster, Lahav \& Fisher 1997; Lahav et al., in preparation). 
This approach is obviously only valid for a uniformly-selected catalogue with 
full sky coverage, which unfortunately is not the case here.
Moreover, clusters and voids outside the `plane' may confuse the 
analysis as well. We prefer instead a more direct approach, 
where each catalogue (${87GB}_{raw}$ and ${PMN}_{raw}$) is analysed independently 
by histograms of number counts.

In order to
explore whether the SGP  is  indeed  visible in the
${87GB}_{raw}$ and ${PMN}_{raw}$ catalogues, we now calculate the surface density of sources as
a function of position in Equatorial ($\alpha$,\,$\delta$), Galactic
($\ell$,\,$b$) and Supergalactic ($\mathit{SGL}$,\,$\mathit{SGB}$) coordinate
systems.  The ${87GB}_{raw}$ and ${PMN}_{raw}$ catalogues are treated here separately, to
avoid being mislead by the flux-density miscalibration highlighted
earlier.
In the study of the SGP we use the more conservative flux limit 
of 100 mJy.

\begin{figure}
\begin{center}
\parbox{8cm}{\psfig{figure=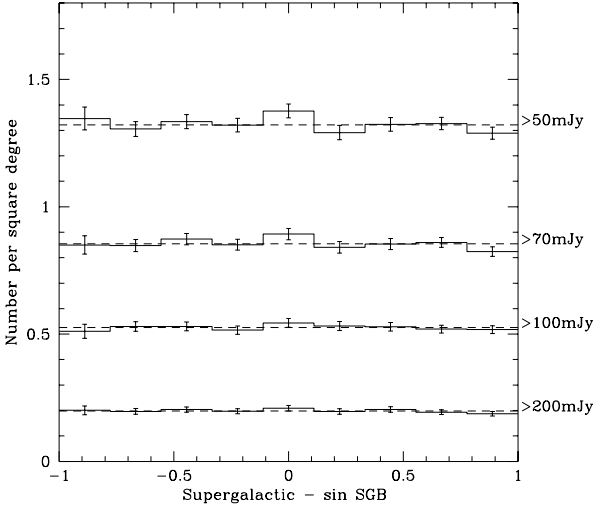,height=8cm}}
\end{center}
  \caption[Surface density of 87GB with $\mathit{SGB}$ and flux-density
   limit]{Source surface density of ${87GB}_{raw}$, as a function of $\mathit{SGB}$
    and flux-density limit.  Error bars are Poissonian.  The 9 bins are
    evenly spaced in $\sin(\mathit{SGB})$.  The area of each bin is
    estimated by Monte Carlo integration.  The dotted line shows the mean
    source density over the whole survey.}
\label{gb-sgb-fig}
\end{figure}

\begin{figure}
\begin{center}
\parbox{8cm}{\psfig{figure=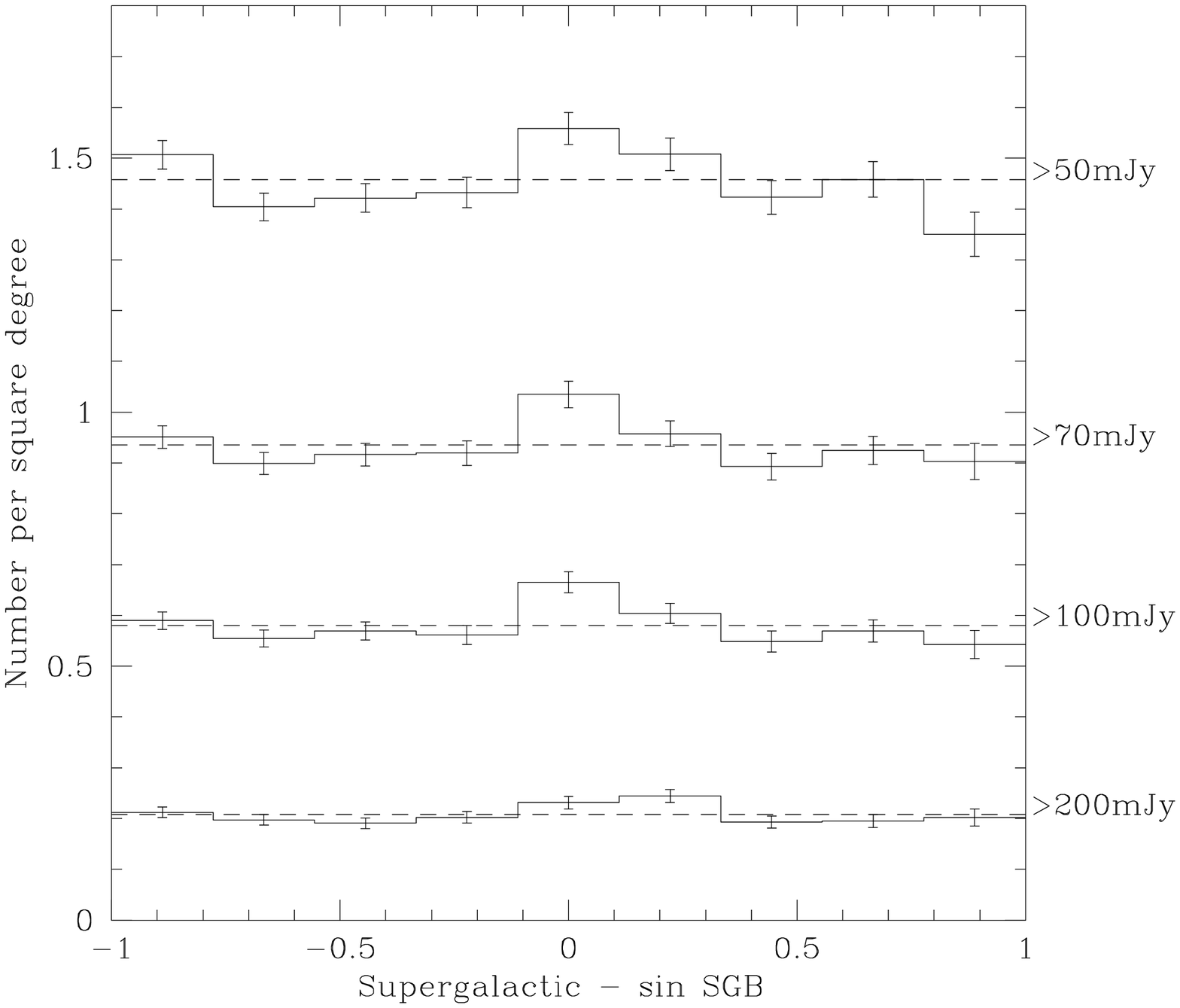,height=8cm}}
\end{center}
  \caption[Surface density of ${PMN}_{raw}$ with $\mathit{SGB}$ and flux-density
  limit]{Source surface density of ${PMN}_{raw}$, as a function of $\mathit{SGB}$
    and flux-density limit.  Error bars are Poissonian.  The 9 bins are
    evenly spaced in $\sin(\mathit{SGB})$.  The area of each bin is
    estimated by Monte Carlo integration.  The dotted line shows the mean
    source density over the whole survey.}
\label{pmn-sgb-fig}
\end{figure}

\begin{figure}
\begin{center}
\parbox{8cm}{\psfig{figure=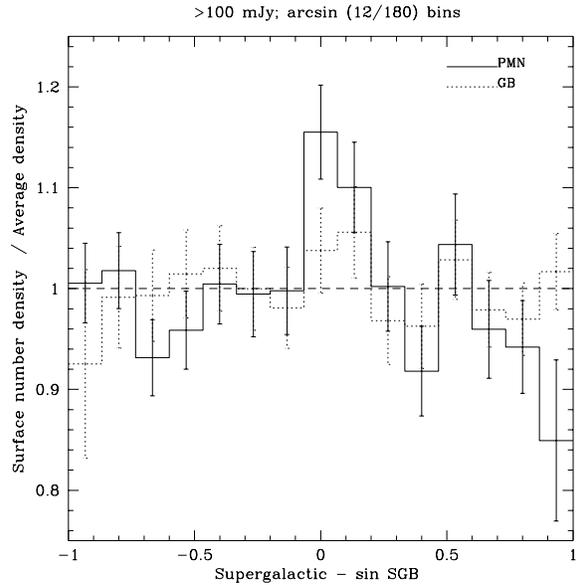,height=8cm}}
\end{center}
  \caption[Surface density of 87GB and PMN above 100\,mJy with
  $\mathit{SGB}$ in 15 bins]{Surface density of sources above 100\,mJy in
    the ${87GB}_{raw}$ and ${PMN}_{raw}$ catalogues, as a function of $\mathit{SGB}$.  Error
    bars are Poissonian.  The 15 bins are evenly spaced in
    $\sin(\mathit{SGB})$.  The area of each bin is estimated by Monte Carlo
    integration.  Results for each catalogue are calculated separately, and
    normalized by the average number density in that catalogue.}
\label{gb-pmn-sgb1-fig}
\end{figure}

In Supergalactic coordinates, the sky is divided into strips of roughly
equal area, equally spaced in $\sin(\mathit{SGB})$ and covering the full
range of $\mathit{SGL}$.  The density of sources in each strip is found by
adding up the number of sources in each catalogue within each strip, and
dividing by the masked area of the strip (that is, the area of the strip
inside the mask for each catalogue).  The total area of a strip may be
easily calculated using spherical geometry, but this simple calculation is
complicated by the numerous parts of the sky that are excluded from the
analysis by the choice of the mask.  Instead, the area of each strip within
the mask is estimated by Monte Carlo integration.  Many points are put down
at random positions within each strip, and the number of points included
within the masked area is calculated.  Thus, the area of each strip is the
ratio of included random points to the total number of random points,
multiplied by the total (unmasked) area of the strip.  This estimate of the
area is found to be accurate to within a small fraction of 1~per~cent.

Figures ~\ref{gb-sgb-fig} and \ref{pmn-sgb-fig} show how the number of
sources in the ${87GB}_{raw}$ and ${PMN}_{raw}$ catalogues changes with the flux-density limit
and as a function of $\mathit{SGB}$.  Both catalogues appear to be complete
above 50\,mJy, but display incompleteness at the 35\,mJy limit (not shown).
The ${87GB}_{raw}$ plot, Figure~\ref{gb-sgb-fig}, is consistent with a uniform
distribution at all flux limits, although there is a suggestion of a small
excess near $\mathit{SGB}=0\degr$.  The ${PMN}_{raw}$ plot, Figure~\ref{pmn-sgb-fig},
displays a strong density enhancement near the Supergalactic equator
($\mathit{SGB}=0\degr$) at all complete flux limits.

\begin{table*}
\begin{minipage}{80mm}
    \caption[Excess counts above 100\,mJy near SGP for 87GB]{\label{radio-gb-tab} Excess number
      of radio sources close to the SGP for ${87GB}_{raw}$, above 100\,mJy.  Total
      number = 8272, Area = 15732\,deg$^2$, mean density =
      0.526\,deg$^{-2}$.  The number expected ($\bar{N}$) is calculated
      from the mean density.  The distance from the mean is the number of
      standard deviations ($\sqrt{N}$) that the detected source excess
      ($N-\bar{N}$) lies from the expected number.  Note that the mean
      density is an over-estimate since it includes any over-density close
      to the SGP.  Thus, the distance from the mean is under-estimated.}
    \begin{tabular}{llllllll} \hline
   Central bin   & Bin area  & \# found & \# expected  & \# excess     & Distance from mean  \\
   width (deg) & (deg$^2$) &  ($N$)   & ($\bar{N}$)  & ($N-\bar{N}$) & ($N-\bar{N}/\sqrt{N}$)\\ \hline
      $\pm$3.82  &  1103     &  602   &   580    &  22    &  0.9 \\
      $\pm$6.38  &  1819     &  989   &   956    &  33    &  1.0 \\
      $\pm$11.54 &  3237     & 1741   &  1702    &  39    &  0.9 \\
      $\pm$19.47 &  5335     & 2828   &  2805    &  23    &  0.4 \\ \hline 
    \end{tabular}
  \end{minipage}
%  \label{radio-gb-tab}
\end{table*}

\begin{table*}
  \begin{minipage}{80mm}
  \caption[Excess counts above 100\,mJy near SGP for PMN]{\label{radio-pmn-tab} Excess number of
    radio sources close to the SGP for ${PMN}_{raw}$, above 100\,mJy.  Total number =
    7829, total area = 13486\,deg$^2$, mean density = 0.581\,deg$^{-2}$.
    }
    \begin{tabular}{llllllll} \hline
   Central bin   & Bin area  & \# found & \# expected  & \# excess     & Distance from mean  \\
   width (deg) & (deg$^2$) &  ($N$)   & ($\bar{N}$)  & ($N-\bar{N}$) & ($N-\bar{N}/\sqrt{N}$)\\ \hline
      $\pm$3.82  &   914     &  613     &   531        &   82          &  3.3 \\
      $\pm$6.38  &  1534     &  1020    &   891        &  129          &  4.1 \\
      $\pm$11.54 &  2766     &  1742    &  1606        &  136          &  3.3 \\
      $\pm$19.47 &  4604     &  2807    &  2673        &  134          &  2.5 \\ \hline 
    \end{tabular}
  \end{minipage}
%  \label{radio-pmn-tab}
\end{table*}

Figure ~\ref{gb-pmn-sgb1-fig} shows how the surface density of sources above
a flux-density limit of 100\,mJy varies as a function of $\mathit{SGB}$ in
the ${87GB}_{raw}$ and ${PMN}_{raw}$ catalogues.  There is a significant increase in source
surface density close to the SGP in the ${PMN}_{raw}$ catalogue, but any increase is
not statistically significant in the ${87GB}_{raw}$ catalogue. Reploting this histogram for a different binning scheme retained this excess around the SGP, indicating that these result are not sensitive to the choice of bin
size.  Tables ~\ref{radio-gb-tab} and ~\ref{radio-pmn-tab} quantify the
excess density within the central strip (either side of $SGP=0\degr$) for
the ${87GB}_{raw}$ and ${PMN}_{raw}$ catalogues, respectively.  Approximately $70 \%$ of
the SGP in each hemisphere is included in the central strip.  Repeating
this analysis with strips of finite width in $\mathit{SGL}$, both for the
whole range of $\mathit{SGB}$ and also for sources within 10$\degr$ of the
SGP, shows that no single cluster of sources causes this increase, but that
the over-density close to the SGP is distributed along the Supergalactic
equator.

A similar procedure was followed in Equatorial coordinates, plotting
similar histograms with bins equally spaced in $\sin(\delta)$ and covering
the whole range of $\alpha$ 
%(see Figure~\ref{radio-surf-fig} and
%Section~\ref{radio-compare-ssec}).  
It is clear that ${87GB}_{raw}$ and ${PMN}_{raw}$ have
slightly different flux calibrations, in accord with the discussion 
earlier. 
Even far above the published flux-density limits of both surveys
($\sim$ 50 mJy) there are more sources per unit area above a given flux
limit in ${PMN}_{raw}$ than there are in ${87GB}_{raw}$.  Fortunately, since both surveys are
analysed separately, this problem will not affect these results.  However,
it does mean that the flux limits given here are somewhat different from
the real flux limits.  This difference will not greatly affect the
generally observed trend, because the radio luminosity function is so
broad.  There is also the possibility of biases caused by ignoring sources
which lie just below the flux limit through experimental error in the
surveys.  Apart from the disturbing flux mismatch highlighted here, no
significant density enhancements are seen in Equatorial coordinates.

No significant patches of increased surface density are seen when a
similar analysis is performed in Galactic coordinates ($\ell$,\,$b$),
plotting histograms of bins equally spaced in $\sin(b)$.  Failure to detect
any increase in source surface density close to the Galactic Plane shows
that there is very little contamination from Galactic sources.  

Similar analyses were conducted, binning the sources in bins of equal width
in right ascension ($\alpha$), Galactic longitude ($\ell$) and
Supergalactic longitude ($\mathit{SGB}$).  Again, no significant departures
from the mean sources density were observed.

\subsection{Comparison with previous studies}

The ${PMN}_{raw}$ survey shows a significant enhancement of the radio source surface
density within $\sim10\degr$ of the Supergalactic Plane: an increase of
$\sim 15 \% $ compared to the mean value.  This is good evidence for
concentration of radio sources towards the SGP, although it is puzzling
that there is no corresponding over-density in the northern sky.  The
concentration of low-Supergalactic-latitude sources in ${PMN}_{raw}$ is not due to one
cluster of sources, but rather it is spread along the whole SGP.  The
number of sources necessary to create this enhancement is quite small (
$\sim$200).

The tests for increased numbers of radio sources close to the SGP follows a
similar analysis of the Molonglo 408\,MHz survey of the southern hemisphere
(Shaver \& Pierre 1989).  Shaver \& Pierre found a significant increase in
the surface density of sources close to the SGP.  This work confirms the
presence of significant concentration of radio sources within 10$\degr$ of
the standard SGP equator 
in the southern hemisphere (${PMN}_{raw}$), but this is not mirrored in the
north (${87GB}_{raw}$).  
 Shaver (1991) pointed out 
that there may be an asymmetry
in the SGL-distribution of the radio galaxies 
(in the Great Attractor direction), 
and this would explain the fact that the SG concentration is more
evident in the southern hemisphere than in the north. 
Shaver \& Pierre (1989) also analysed a small 
redshift survey, finding that there was a significant preference for nearby
($z<0.02$) radio sources to lie close to the SGP, but that this preference
was not observed for radio sources at higher redshift ($0.02<z<0.1$).

The present analysis finds an over-density of about 130 radio sources with a
4.85\,GHz flux density above 100\,mJy within an area of about 1500 square
degrees, up to $10\degr$ either side of the SGP in the southern hemisphere.
A less significant over-density is also observed in the north.  The models
of Dunlop \& Peacock (1990) 
predict that the surface density of radio sources
above 100\,mJy within a redshift $z<0.02$ is $\approx$18 
per str
%$\psr$
(1\,str=3282.97\,square degrees).  It is interesting to note that the
over-density found along the SGP in the ${PMN}_{raw}$ survey could be accounted for by
approximately the same number of radio sources found within a redshift
$z<0.02$ ($\sim 60 h^{-1}$Mpc).

If we assume that the over-density in the ${PMN}_{raw}$ survey is caused by nearby
radio galaxies out to a redshift $z=0.02$, then we can estimate the
thickness of SGP.  The angle subtended by the over-dense strip is
$\theta\sim20\degr$, and thus, at the distance corresponding to $z=0.02$
($d= cz/H_0 =60 h^{-1}$Mpc), the SGP has a thickness $\sim
2\,d\tan(\theta/2)\approx20 h^{-1}$Mpc.  This also raises the question of how
far the SGP extends.  One test of the hypothesis that the over-density
close to the SGP really is caused by nearby radio sources would be to
cross-correlate the radio sources with those in the optical 
IRAS catalogues, 
in a similar way to the study by Shaver (1991).

It may be possible to cross-identify objects that are both optical IRAS sources
and also radio sources in the ${87GB}_{raw}$   and ${PMN}_{raw}$  catalogues.  Repeating this
analysis after removing radio sources that appear in the optical 
and IRAS catalogues
may show whether the apparent over-density close to the SGP is caused by a
flattened distribution of nearby sources.

It could be argued that these results may be pure statistical fluke, and
that the detected over-density is caused by a relatively small number of
radio sources.  However, it is not clear how a bias towards detecting
sources that lie close to the SGP could have entered the radio surveys.

On large angular scales, there is a clear detection of the SGP in the
southern sky, but not in the north.  The over-density in the south may be
attributed to nearby radio galaxies lying preferentially toward the SGP,
but it is not clear why there is no significant detection in the north.
Possible explanations of this conundrum include the possibilities that
nearby sources are somehow excluded from ${87GB}_{raw}$ , or that they are counted
more than once in ${PMN}_{raw}$  (\eg nearby double-lobed radio sources may be
included twice).  Cross-identification of sources in the radio surveys with
optical and IRAS galaxies  (both relatively shallow surveys, out to
$\approx 6000$ km/s) could be used to find the nearby radio sources, to
check if they are indeed responsible for the over-density close to the SGP.

\section{Discussion}

We have studied in this paper the possibility of using 
radio surveys 
with median redshift $z \sim 1$ to probe large scale structure.
In particular, we have utilised the technique of spherical harmonics
to predict 
the amplitude of the dipole and higher
in the angular distribution of radio galaxies.

Our conclusions are:

(a)   The dipole is due to 2 effects which turn out to be of
comparable magnitude: (i) our motion with respect to the 
CMB, and (ii) large scale structure,

(b) Catalogues like 87GB and PMN have the potential of probing 
structure on large scales, e.g.  
The quadrupole measures scales 
of $k^{-1} \sim 600 h^{-1}$ Mpc. 

(c) Unfortunately, the Poisson shot noise in these sparse catalogues
    is expected to be large than the clustering signals, as predicted 
    for a family of Cold Dark Matter models.

(d)   However, we detect dipole and higher harmonics in the combined ${87GB-PMN}_{raw}$ 
  catalogue which are far larger than expected.  We
  attribute this to a 2\% 
  flux mismatch between the two catalogues. 
  Ad-hoc corrections to match the catalogues may suggest a marginal
  detection of a dipole.

There are two likely explanations for this. One is
that we are still seeing the effect of survey geometry, or incompatibility
of data reduction algorithms in producing the catalogues.
The second is that, because the catalogues represent only one realization of
the ensemble average harmonics we predict, we may be detecting local
structure that causes the large amplitude we measure.

(e)   To detect a dipole and higher harmonics unambiguously,
  a catalogue with full sky coverage and $\sim 10^6$ sources is
  required.  

(f)   We have investigated the existence and extent of the 
   Supergalactic Plane in 
  the above catalogues. In a strip of $\pm 10^\circ$ 
  of the standard Supergalactic equator  
  we find a $3\sigma$ detection in ${PMN}_{raw}$, 
  but only $1\sigma$ in ${87GB}_{raw}$.
This analysis demonstrates that large-scale structure studies based on new
radio surveys nearing completion will provide important estimates of distant
structure and its evolution.
On-going surveys with the VLA, FIRST (Becker, White \& Helfand 1995) and
NVSS (Condon 1997), will yield approximately $10^6$ sources over
the sky, thus overcoming the limitations imposed by the shot-noise
at the  current 4.85\,GHz surveys.  
Furthermore, the positional accuracy of these
new surveys will by sufficient for optical spectroscopy using candidates
directly from the catalogues.  Although difficult to obtain,
redshifts to the radio sources are essential 
in order to improve the measurement of the power-spectrum 
by localizing the clustering pattern in 3 dimensions.

\bigskip
\bigskip

{\bf Acknowledgements:} 
We thank A. Fabian, E. Gawiser, M. Hudson, 
A. Heavens, A. Lasenby, T. Piran, M. Rees, C. Scharf, P. Shaver, 
M. Treyer, M. Webster and K. Wu for helpful discussions and comments.
AB was supported by a Winston Churchill Scholarship and a NSF Graduate Fellowship. AJL acknowledges the receipt of a PPARC studentship.

\end{document}